\documentclass[12pt,amsbsy]{article}
\newcommand{\beq}{\begin{equation}}
\newcommand{\eeq}{\end{equation}}
\newcommand{\beqn}{\begin{eqnarray}}
\newcommand{\eeqn}{\end{eqnarray}}

\newcommand{\ga}{\mbox{${\gamma}$}}

\begin{document}

\begin{center}
{\bf \large Holographic bound and spectrum of quantized black hole}
\end{center}

\begin{center}
I.B. Khriplovich\footnote{khriplovich@inp.nsk.su}
\end{center}
\begin{center}
Budker Institute of Nuclear Physics\\
630090 Novosibirsk, Russia,\\
and Novosibirsk University
\end{center}

\bigskip

\begin{abstract}
Recent attempts to calculate the black-hole entropy in loop quantum
gravity are demonstrated to be erroneous. The correct solution of
the problem is pointed out.
\end{abstract}

\hspace{3mm} PACS numbers: 04.60Pp, 04.60.Ds, 04.60.Nc, 03.65.Sq

\vspace{8mm}

The description of a quantized surface in loop quantum gravity (LQG)
looks as follows (see, e.g. \cite{ash}). One ascribes to this
surface a set of punctures, each of them supplied with an integer or
half-integer quantum number $j$:
\beq\label{j}
j= 1/2, 1, 3/2, ...\; .
\eeq
This ``angular momentum'' $j$ has as usual $2j+1$ projections $m$:
\beq\label{m}
m = -j\,,\,...\, j\,.
\eeq
The area of a surface in LQG is
\beq\label{Aj}
A =8\pi\ga\, l_p^2 \sum_i \sqrt{j_i(j_i+1)}\,.
\eeq
The numerical factor $\ga$ in (\ref{Aj}) cannot be determined
without an additional physical input. This ambiguity originates from
a free (so-called Barbero-Immirzi) parameter \cite{imm,rot} which
corresponds to a family of inequivalent quantum theories, all of
them being viable without such an input. Once the general structure
of the horizon area in LQG is fixed, the problem is to determine
this overall factor, i.e. the Barbero-Immirzi (BI) parameter.

First attempts to fix the value of this parameter by calculating the
black hole entropy~\cite{rov,kra} did not lead to concrete
quantitative results.

Then it was argued in~\cite{asht} that for the area of a black hole
horizon one obtains all $j=1/2$. With two possible projections $\pm
1/2$ for the quantum number $j$, the total number $K$ of degenerate
quantum states of the horizon is
\beq\label{kifb}
K=2^\nu.
\eeq
Correspondingly, its entropy is
\beq\label{S1/2}
S_{1/2}= \ln K= \nu \ln 2.
\eeq
From it, with the Bekenstein-Hawking relation
\beq\label{BH}
S = \frac{A}{4 l_p^2}\,,
\eeq
one obtains the following value for the BI parameter:
\beq\label{ga1}
\ga = \frac{\ln 2}{\pi \sqrt{3}}\,=0.127.
\eeq

However, this result of \cite{asht} was demonstrated in~\cite{kh} to
be certainly incorrect since it violates the holographic bound
formulated in ~[8-10].

Later, the result of \cite{asht} was also criticized and revised in
\cite{lew,mei}, with the conclusion that the number of the relevant
horizon states is underestimated in \cite{asht}. The value of the BI
parameter, as calculated in~\cite{mei}, is
\beq\label{ga2}
\ga = 0.238.
\eeq

In the present note I demonstrate that the results of~\cite{lew,mei}
are also in conflict with the holographic bound. Besides, an
apparent error in state counting made in~\cite{lew,mei} is pointed
out.

But first of all on the holographic bound itself. According to it,
the entropy $S$ of any spherically symmetric system confined inside
a sphere of area $A$ is bounded as follows:
\beq\label{hb}
S \leq \frac{A}{4 l_p^2}\,,
\eeq
with the equality attained only for a system which is a black hole.
This result can be formulated otherwise. Among the spherical
surfaces of a given area, it is the surface of a black hole horizon
that has the largest entropy.

A simple intuitive argument confirming bound (\ref{hb}) is as
follows~\cite{sus}. Let us allow the discussed system to collapse
into a black hole. During the collapse the entropy increases from
$S$ to $S_{bh}$, and the resulting horizon area $A_{bh}$ is
certainly smaller than the initial confining one $A$. Now, with the
account for the Bekenstein-Hawking relation (\ref{BH}) for a black
hole, we arrive, through the obvious chain of (in)equalities:
\[
S \leq S_{bh} = \frac{A_{bh}}{4 l_p^2} \leq \frac{A}{4 l_p^2}\,,
\]
at the discussed bound (\ref{hb}).

In fact, the correct value of the BI parameter $\ga$ was obtained by
us few years ago~\cite{kk}(see also~\cite{kh,kh1}). Our derivation
was based exactly on the requirement that the surface of a black
hole horizon should have maximum entropy.

To substantiate our objections to the results
of~\cite{asht,lew,mei}, I sketch below the derivation of our result
(for more technical details see~\cite{kk}).

We consider the ``microcanonical'' entropy $S$ of a surface (though
with a fixed area, instead of fixed energy). The entropy is defined
as the logarithm of the number of states of this surface with a
fixed area $A$.

To analyze the problem, it is convenient to rewrite formula
(\ref{Aj}) as follows:
\beq\label{Anu}
A = 8\pi\ga\, l_p^2 \sum_{jm}\sqrt{j(j+1)}\;\nu_{jm}\,.
\eeq
Here $\nu_{jm}$ is the number of punctures with given $j$ and $m$.
It can be demonstrated~\cite{kh,kh1} that the only reasonable
assumption on the distinguishability of punctures that may result in
acceptable physical predictions, complying both with the
Bekenstein-Hawking relation and with the holographic bound, is as
follows:\\

\begin{tabular}[h]{cccc}
 \vspace{3mm}
 nonequal $j$, & any $m$ & $\longrightarrow$ &
distinguishable;\\
\vspace{3mm} equal $j$, & nonequal $m$ &
$\longrightarrow$ & distinguishable;\\
 \vspace{3mm}
equal $j$, & equal $m$ & $\longrightarrow$ & indistinguishable.\\
\end{tabular}\\

\noindent Under this assumption, the number of states of the horizon
surface for a given $\nu_{jm}$, is obviously
\beq\label{mk}
K = \nu\,!\, \prod_{jm}\,\frac{1}{\nu_{jm}\,!}\;, \quad \nu =\sum_j
\nu_j = \sum_{jm} \nu_{jm}\,,
\eeq
and the corresponding entropy equals
\beq\label{ms}
S=\ln K = \ln(\nu\,!)\,- \sum_{jm}\,\ln(\nu_{jm}\,!)\,.
\eeq
The structures of the last expression and of formula (\ref{Anu}) are
so different that in a general case the entropy certainly cannot be
proportional to the area. However, this is the case for the maximum
entropy in the classical limit.

In the classical limit, with all effective ``occupation numbers''
large, $\nu_{jm} \gg 1$, the entropy in the Stirling approximation
is
\beq\label{en2}
S= \nu \ln \nu -\sum_{jm} \nu_{jm} \ln \nu_{jm}\,.
\eeq
We calculate its maximum value for a fixed area $A$, i.e. for a
fixed sum
\beq\label{N}
N\,=  \sum_{jm}\sqrt{j(j+1)}\,\nu_{jm}={\rm const}.
\eeq

The problem reduces to the solution of the system of equations
\beq\label{sys}
\ln \nu - \ln \nu_{jm} = \mu \sqrt{j(j+1)}\,,
\eeq
where $\mu$ is the Lagrange multiplier for the constraining relation
(\ref{N}). These equations can be rewritten as
\beq\label{nu1}
\nu_{jm}=\nu \,e^{- \mu \sqrt{j(j+1)}}\,,
\eeq
or
\beq\label{nu2}
\nu_j = (2j+1)\,e^{- \mu \sqrt{j(j+1)}}\,\nu \,.
\eeq
Now we sum expressions (\ref{nu2}) over $j$, and with $\sum_j \nu_j
= \nu$ arrive at the equation for $\mu$
\beq\label{equ}
\sum_{j=1/2}^{\infty} (2j+1)\, e^{- \mu \sqrt{j(j+1)}} = 1\,,
\eeq
with the solution
\beq\label{mu}
\mu = 1.722.
\eeq

On the other hand, when multiplying equation (\ref{sys}) by
$\nu_{jm}$ and summing over $jm$, we arrive with the constraint
(\ref{N}) at the following result for the maximum entropy for a
given value of $N$:
\beq\label{en}
S_{\rm max}= 1.722\,N\,,
\eeq
so that with the Bekenstein-Hawking relation and formula (\ref{Anu})
we find the value of the BI parameter\footnote{Quite recently this
calculation with the same result, though with somewhat different
motivation, was reproduced in~\cite{gm}.}:
\beq\label{bip}
\ga = 0.274.
\eeq

Let us come back now to the results (\ref{ga1}) and (\ref{ga2}). If
one agrees that the BI parameter is universal (i.e. its value is not
special to black holes, but refers to any quantized spherical
surface), then it is self-obvious that the results
of~\cite{asht,lew,mei} are in conflict with the holographic bound
and therefore incorrect. Indeed, with relations (\ref{m}) and
(\ref{Aj}) valid for a generic quantized spherical surface (be it an
horizon or not) this is our solution (\ref{bip}) that describes a
black hole since just it corresponds to the maximum entropy.

The error made in~\cite{asht} has been acknowledged
recently~\cite{asht1}, so we do not discuss anymore the result
(\ref{ga1}). However, it would be useful to indicate an apparent
error in state counting made in~\cite{lew,mei}. It can be easily
checked that the transition from formula (25) to formulae (29), (36)
of~\cite{lew} performed therein and then employed in~\cite{mei}, is
certainly valid under the assumption that for each quantum number
$j$ only two maximum projections $\pm j$ are allowed, instead of
$2j+1$. No wonder therefore that the equation for the BI parameter
in~\cite{mei} looks as
\beq\label{2}
2\sum_{j=1/2}^{\infty}\, e^{- \mu \sqrt{j(j+1)}} = 1\,,
\eeq
instead of ours (\ref{equ}) (see also the discussion of (\ref{2})
in~\cite{gm}).

The conclusion is obvious. Any restriction on the number of
admissible states for the horizon, as compared to a generic
quantized surface, be it the restriction to
\[
j=1/2\,, \quad m=\pm 1/2\,,
\]
made in~\cite{asht}, or the restriction to
\[
{\rm any} j\,, \quad m=\pm j\,,
\]
made in~\cite{lew,mei}, results in a conflict with the holographic
bound.

\newpage

\begin{center}***\end{center}

I truly appreciate the kind hospitality extended to me at School of
Physics, University of New South Wales, Sydney, where the major part
of this work was done. I am sincerely grateful to O.P. Sushkov for
the warm invitation to UNSW and for useful discussions.

The investigation was supported in part by the Russian Foundation
for Basic Research through Grant No.03-02-17612.


\begin{thebibliography}{99}
\bibitem{ash} Ashtekar A and Lewandowski J 1997 {\em Class. Quantum Grav.} {\bf 14}
55 \\ ({\em Preprint} gr-qc/9602046)
\bibitem{imm} Immirzi G 1997 {\it Class. Quantum Grav.} {\bf 14} L177 ({\em Preprint}
gr-qc/9701052)
\bibitem{rot} Rovelli C and Thiemann T 1998 {\it Phys. Rev.} {\bf D57} 1009 ({\em Preprint}
gr-qc/970559)
\bibitem{rov} Rovelli C 1996 {\it Phys. Rev. Lett.} {\bf 77} 3288 ({\em Preprint}
gr-qc/9603063)
\bibitem{kra} Krasnov K V 1997 {\it Phys. Rev.} {\bf D55} 3505 ({\em Preprint} gr-qc/9603025);\\
Krasnov K V 1998 {\it Gen. Rel. Grav.} {\bf 30} 53 ({\em Preprint}
gr-qc/9605047);\\ Krasnov K V 1999 {\it Class. Quantum Grav.} {\bf
16} 563 ({\em Preprint} gr-qc/9710006)
\bibitem{asht} Ashtekar A, Baez J, Corichi A and Krasnov K 1998
{\it Phys. Rev. Lett.} {\bf 80} 904 \\ ({\em Preprint}
gr-qc/9710007)
\bibitem{kh} Khriplovich I B 2004 {\it Zh. Eksp. Teor. Fiz.}
{\bf 126} 527 \newline [{\it  Sov. Phys. JETP}$\,$ {\bf 99} 460 ]
({\em Preprint} gr-qc/0404083)
\bibitem{bek} Bekenstein J D 1981 {\it Phys. Rev.} {\bf D23} 287
\bibitem{tho} 't Hooft G 1993 in {\it Salam Festschrift} Singapore
({\em Preprint} gr-qc/9310026)
\bibitem{sus} Susskind L 1995 {\it J. Math. Phys.} {\bf 36} 6377 ({\em Preprint}
gr-qc/9710007)
\bibitem{lew} Lewandowski J, Domagala M 2004 {\it Class. Quantum Grav.} {\bf
21} 5233\\ ({\em Preprint} gr-qc/0407051)
\bibitem{mei} Meissner K 2004 {\it Class. Quantum Grav.} {\bf
21} 5245 ({\em Preprint} gr-qc/0407052)
\bibitem{kk} Korkin R V, Khriplovich I B 2002 {\it Zh. Eksp. Teor. Fiz.} {\bf
122} 1\newline [{\it Sov. Phys. JETP}$\,$ {\bf 95} 1] ({\em
Preprint} gr-qc/0112074)
\bibitem{kh1} Khriplovich I B 2004 ({\em Preprint} gr-qc//0409031)
\bibitem{gm} Ghosh A, Mitra P 2004 ({\em Preprint} gr-qc//0411035)
\bibitem{asht1} Ashtekar A 2004 ({\em Preprint} gr-qc/0410054)

\end{thebibliography}
\end{document}